\documentclass[aps,prd,secnumarabic,amssymb, amsmath,nobibnotes,nofootinbib,11pt,superscriptaddress]{revtex4}
\usepackage{amsfonts,amsmath,hyperref,url, color}
\usepackage{bm, bbm}
\usepackage{graphicx}
\usepackage{mathtools}
\usepackage{mathrsfs}

\begin{document}
\title{$\kappa$-deformed BMS symmetry}

\author{Andrzej Borowiec}
\affiliation{Institute for Theoretical Physics, University of Wroc\l{}aw, pl.\ M.\ Borna 9, 50-204 Wroc\l{}aw, Poland}

\author{Lennart Brocki}
\affiliation{Institute for Theoretical Physics, University of Wroc\l{}aw, pl.\ M.\ Borna 9, 50-204 Wroc\l{}aw, Poland}

\author{Jerzy Kowalski-Glikman}
\affiliation{Institute for Theoretical Physics, University of Wroc\l{}aw, pl.\ M.\ Borna 9, 50-204 Wroc\l{}aw, Poland}
\affiliation{National Centre for Nuclear Research, Ho\.{z}a 69, 00-681 Warsaw, Poland}

\author{Josua Unger}
\affiliation{Institute for Theoretical Physics, University of Wroc\l{}aw, pl.\ M.\ Borna 9, 50-204 Wroc\l{}aw, Poland}

\date{\today}

\begin{abstract}
We present the quantum $\kappa$-deformation of BMS symmetry, by generalizing the lightlike $\kappa$-Poincar\'e Hopf algebra. On the technical level our analysis relies on the fact that  the lightlike $\kappa$-deformation of Poincar\'e algebra is given by a twist and the lightlike deformation of any algebra containing Poincar\'e as a subalgebra can be done with the help of the same twisting element. We briefly comment on the physical relevance of the obtained $\kappa$-BMS Hopf algebra as a possible asymptotic symmetry of quantum gravity.
\end{abstract}

\maketitle

In the recent years there is a surge of interest in BMS symmetry at null infinity of asymptotically flat spacetimes \cite{Bondi:1962px}, \cite{Sachs:1962wk}, \cite{Sachs:1962zza}. This renewed interest in the seemingly exotic aspects of classical general relativity was fueled by the discovery of a surprising and close relationship between asymptotic symmetries and soft gravitons theorem \cite{Weinberg:1965nx} that, as it turned out, has its roots in Ward identities for supertranslations \cite{Strominger:2013jfa}, \cite{He:2014laa}. Moreover, the gravitational memory effect \cite{Christodoulou:1991cr} appears to be related to the two \cite{Strominger:2014pwa}, so that there is a triangle of interrelationships, which vertices are BMS symmetry, Weinberg's soft graviton theorem, and the memory effect. The extensive discussion of these effects can be found in recent reviews \cite{Strominger:2017zoo} and \cite{Compere:2018aar}.

It has also been argued recently \cite{Hawking:2016msc} that the similarity of null infinity and black hole horizon suggests that the charges associated with the BMS symmetry might be present at the black hole horizon, so that the black hole might have an infinite number of hairs, so-called soft hairs, in addition to the three classical one: mass, charge, and angular momentum. It was argued in \cite{Hawking:2016msc} and \cite{Hawking:2016sgy} that the presence of these charges may help solving the black hole information paradox.

In this letter we will investigate the properties of a $\kappa$-deformed generalization of the BMS symmetry. There are several reasons to be interested in such a generalization. The main one is that it is believed that investigations of $\kappa$-deformation might shed some light on the properties of elusive quantum gravity. The deformation parameter of $\kappa$-deformation of Poincar\'e algebra   \cite{Lukierski:1991pn}--\cite{kappaM2}, see \cite{Kowalski-Glikman:2017ifs} for a recent review and more references, has dimension of mass, and therefore it is natural to identify it with Planck mass, or inverse Planck length, which in turn suggests a possible close relationship between this deformation and quantum gravity.
In fact, it was shown in  \cite{Cianfrani:2016ogm} that in the case of gravity in $2+1$ dimensions the $\kappa$-Poincar\'e algebra is the algebra of symmetries of quantum flat space in non-perturbatively quantized quantum gravity, and therefore it plays in quantum gravity the role analogous to the fundamental role played by the Poincar\'e algebra in quantum field theory.

There are three incarnations in which $\kappa$-deformation can appear. First, there is the mostly studied deformation in time direction, in which case the rotations subalgebra of the Poincar\'e algebra is kept undeformed. This deformation satisfies a natural requirement that  rotations are not deformed, and in particular that the angular momentum and spin composition laws are not modified.  It is for this same reason that space-like $\kappa$-deformation with a preferred spacial direction of deformation was never considered to be a phenomenologically viable possibility, and is treated as a mathematical curiosity with no real relevance for physics.

The third possibility is the lightlike $\kappa$-deformation \cite{Ballesteros:1995mi} -- \cite{AP15}. In contrast to the previous two cases it is implemented by a triangular $r$-matrix, which satisfies the Classical Yang-Baxter Equation.  A remarkable property of this so-called non-standard deformation is that it is given by a twist which makes its structure much simpler to construct \cite{Drinfeld}. Particularly, any embedding of the algebra into a larger Lie algebra preserves the triangular structure and allows to perform the deformation by the same twist element. Since the BMS algebra is ultimately tied in with null infinity and contains the Poincar\'e subalgebra, in this paper we deform it using the lightlike $\kappa$-deformation. Another characteristic  property of the twist quantization is that it deforms the coalgebraic sector of the corresponding universal enveloping algebra, leaving the Lie algebra untouched.\footnote{However, it should be noticed that the first explicit formulas for the light-like $\kappa$-deformed coproducts have been provided in a non-classical basis  \cite{Ballesteros:1995mi,LMM,Mudrov} without using the twist.}
Moreover, knowing the twist one can also obtain the underlying noncommutative quantum-deformed spacetime structure.

Let us start reviewing the construction of the lightlike deformed Poincar\'e algebra as a Hopf algebra following \cite{Borowiec:2013lca}. Then we will extend this procedure to the BMS algebra. We start with the Poincar\'e algebra in four dimensions with light-cone generators
\begin{align}
M_{+-},\  M_{+i}, \ M_{-i},\  M_{ij}, \ P_+,\  P_-,\ P_i\,,\quad i,j = 1,2
\end{align}
satisfying
\begin{align}
\left[ M_{+\,i}\,,M_{-\,j}\right]  &=-i\left( M_{ij}+g_{ij}M_{+\,-}\right)
\,, &  \left[ M_{\pm \,i}\,,M_{\pm \,j}\right] &=0  \label{M+aM-b} \\
\left[ M_{\pm \,i}\,,M_{j\,k}\right]  &=i\left( g_{ij}M_{\pm
\,k}-g_{i\,k}M_{\pm \,j}\right) \,, &  \left[ M_{+\,-}\,,M_{\pm \,i}\right]
&=\pm iM_{\pm \,i}\   \label{MpmMbc} \\
\left[ M_{+\,-},P_{\pm }\right]  &=\pm i\,P_{\pm }\,,
 & \left[ M_{\pm \,i}\,,P_{j}\right] &=ig_{ij}P_{\pm }\   \label{M_pmP} \\
\left[ M_{\pm \,i}\,,P_{\pm }\right]  &=\left[ M_{+\,-}\,,P_{i}\right]
=0\,, &  \left[ M_{\pm \,i}\,,P_{\mp }\right] &=-\,iP_{i}
\label{M_pmP_pm}\\
\left[ M_{i\,j}\,,P_{k}\right] &=i\left(g_{ik}P_{j}-g_{jk}P_{i}\right)&&\label{MijPk}
\end{align}
The  metric used here has the nonzero entries $g_{+-} = g_{-+}= 1,\; g_{11} = g_{22} = -1$.

The primitive coproduct structure
\begin{align}
\Delta_0 (X) =  \mathbbm1 \otimes X + X \otimes  \mathbbm1
\end{align}
and antipode
\begin{align}
S_0 (X) = -X,
\end{align}
where $X$ denotes an arbitrary element of the algebra above, makes it a Hopf algebra. This Hopf algebra (established, in fact, on the enveloping algebra) can be deformed by performing a "twist" on the coproduct and the antipode (see below) while leaving the commutator structure unchanged.
\newline

Taking linear transformations
\begin{align}
M^{+-}& = i(l_0 + \bar l_0), \quad M^{12} = (l_0- \bar l_0) \\
M^{+1}& = \frac{i}{\sqrt{2}}(l_{-1} + \bar l_{-1}), \quad M^{+2}= \frac{1}{\sqrt{2}}(l_{-1} - \bar l_{-1})\nonumber \\
M^{-1} & = \frac{i}{\sqrt{2}}(l_{1} + \bar l_{1}), \quad M^{-2} =\frac{-1}{\sqrt{2}}(l_{1} - \bar l_{1})\nonumber \\
P_+ & = \frac{i}{\sqrt{2}}T_{11}, \quad P_- = - \frac{i}{\sqrt{2}} T_{00} \nonumber\\
P_1 & = \frac{i}{2} (T_{10} + T_{01}), \quad P_2 = \frac{-1}{2}(T_{10} - T_{01})\nonumber
\end{align}
one establishes that the Poincar\'e algebra is a subalgebra of the (extended) BMS algebra with generators
$l_m$, $ \bar l_m$, $ T_{pq}$
with $m, p, q \in \mathbb{Z}$, satisfying \cite{Barnich:2009se}, \cite{Barnich:2011mi}
\begin{align}
[l_m, l_n] = (m-n)l_{m+n}, \quad [\bar l_m, \bar l_n ] = (m-n) \bar l_{m+n}, \quad [l_m, \bar l_n]= 0 \label{bmsl} \\
[l_l, T_{m,n}] = \left(\frac{l+1}{2} -m\right)T_{m+l, n}, \quad [\bar l_l, T_{m,n}] = \left(\frac{l+1}{2} -n\right)T_{m, n+l}.\label{bmsT}
\end{align}
This is the infinite dimensional Lie algebra \eqref{bmsl}, \eqref{bmsT} which we are going to deform as a Hopf algebra. It turns out however that this form of the algebra is not the one convenient for our purposes and therefore we will use another, obtained by linear transformations of the generators $l_m$, $ \bar l_m$, $ T_{pq}$. To this end we will use the original Poincar\'e generators $M$ and $P$  satisfying the algebra \eqref{M+aM-b}--\eqref{MijPk} along with the linear combinations of the higher BMS generators defined as follows
\begin{align}
k_n & = l_n + \bar l_n , \quad &\bar k_n  &= -i (l_n - \bar l_n) ,\\
S_{mn}& = \frac{1}{2} (T_{mn} + T_{nm}) , &\quad A_{mn} &= -\frac{i}{2} (T_{mn} - T_{nm})\,.
\end{align}
Notice that we have $k_0=-i M^{+-}$, $\bar k_0 = -i M^{12}$, $k_{1} = -i\sqrt{2} M^{-1}$, $k_{-1} = -i\sqrt{2} M^{+2}$, $\bar k_{1} = -i\sqrt{2} M^{-2}$, $\bar k_{-1} = -i\sqrt{2} M^{+1}$, $S_{00}= i\sqrt2 P_-$, $S_{11} =-i\sqrt{2} P_+$, $S_{01}=-iP_1$, $A_{01}=-iP_2$. In terms of these generators the BMS algebra reads
\begin{align}
[k_n, k_m ] & = (n-m) k_{n+m} , \quad [\bar k_n,\bar k_m ] = -(n-m) k_{n+m},\\
[k_n, \bar k_m ] & = (n-m) \bar k_{n+m} ,\\
[k_n, S_{pq}] & = \left(\frac{n+1}{2}-p \right) S_{p+n, q} + \left( \frac{n+1}{2}-q \right) S_{q+n, p}, \\
[\bar k_n, S_{pq}]& = \left(\frac{n+1}{2}-p \right) A_{p+n, q} + \left( \frac{n+1}{2}-q \right) A_{q+n, p}, \\
[k_n, A_{pq}]& = \left(\frac{n+1}{2}-p \right) A_{p+n, q} + \left( \frac{n+1}{2}-q \right) A_{p, q+n}, \\
[\bar k_n, A_{pq}] & = \left(\frac{n+1}{2}-p \right) S_{p+n, q} + \left( \frac{n+1}{2}-q \right) S_{q+n, p} ,
\end{align}
and one can check that it reproduces the Poincar\'e algebra \eqref{M+aM-b}--\eqref{MijPk} in the sector $|m|, |n|, \ldots\leq1$.

Knowing the undeformed BMS algebra, given by the commutators above with trivial Hopf structure, we can turn to the deformation. To this end we use the fact that the lightlike $\kappa$-deformation of Poincar\'e algebra corresponds to the triangular classical $r$-matrix
$r_{LC}=M_{+\,-}\wedge P_{+}+M_{+\,a}\wedge P^{a}$ found in \cite{Zakrzewski}, for which the twisting elemen $\mathcal{F}$ is known as the so-called extended Jordanian twist proposed in \cite{Kulish:1998be}. It is based on the subalgebra containing six out of ten Poincar\'e generators:  $(M_{+-}, P_+, M_{+1}, P_1, M_{+2}, P_2)$ and has the following form \footnote{From this one also immediately gets the quantum $R$-matrix $\mathcal R= \mathcal{F}_{21}\mathcal{F}^{-1}$.}
\begin{align}
\mathcal{F} & = \exp \left(-i\frac{1}{\kappa} M_{+i} \otimes P^i \right) \exp (-i M_{+-} \otimes \log \Pi_+)\nonumber  \\
&= \exp\left(\frac{1}{\sqrt{2}\kappa}(i k_1 \otimes S_{01} +i \bar k_1 \otimes A_{01})\right)\exp\left( - k_0 \otimes \sum_{j=0}^{\infty} \frac{1}{j+1} \left(\frac{-i S_{11}}{\sqrt{2}\kappa}\right)^{j+1}\right),\label{TwiEl}
\end{align}
where $\Pi_+ = 1 + \frac{P_+}{\kappa}= 1 + \frac{iS_{11}}{\sqrt{2}\kappa}$.

We recall \cite{Drinfeld}  that the twisting element $\mathcal{F} \equiv a_\alpha \otimes b^\alpha \in \mathcal{A} \otimes \mathcal{A}$ has to be an invertible, normalized  2-cocycle for the original (i.e. undeformed) algebra $\mathcal{A}$, i.e.\ it has to fulfill
\begin{align} \label{twist}
\mathcal{F}_{12} \cdot (\Delta_0 \otimes \mathbbm1) (\mathcal{F}) = \mathcal{F}_{23} \cdot (\mathbbm1 \otimes \Delta_0 )(\mathcal{F}), \quad \epsilon(a_\alpha)b^\alpha=1
\end{align}
with $\mathcal{F}_{12} = a_\alpha \otimes b^\alpha \otimes \mathbbm1$, etc. It provides the deformed coproduct via  the similarity transformation $\Delta(X)=\mathcal{F}\Delta_0(X)\mathcal{F}^{-1}$.

The twist deformation can be extended from the Poincar\'e subalgebra to the whole BMS algebra by making repeated use of the Hadamar  formula
$$
e^A B e^{-A} = \sum_{n=0}^{\infty} \frac{1}{n!} \underbrace{ [A,[A, ...[A,}_{\text{n times}} B]..].
$$
As a result, we get the expressions for the coproduct in the superrotation sector \footnote{To simplify these expressions, in the main text we present the expressions to the leading order in $1/\kappa$ only. The complete expressions can be found in the appendix. We also use the standard notation for the Poincar\'e generators on the right hand side of the formulae below, so as to distinguish them from the superrotations and supertranslations. }
\begin{align}
\Delta (k_m) =&  \mathbbm1 \otimes k_m + k_m \otimes \left( \mathbbm1 +
m\frac{P_+}{\kappa} \right)  + \frac{m-1}{\sqrt{2}\kappa}M^{+-} \otimes  S_{1+m,1} \nonumber\\
&+ \frac{i}{\kappa} \left((M^{-1} \otimes \left(\frac{m+1}{2}S_{m1} + \frac{m-1}{2} S_{0,m+1} \right) - M^{-2} \otimes \left( \frac{m+1}{2} A_{m1} - \frac{m-1}{2}A_{0,m+1} \right)  \right) \nonumber \\
&+ \frac{i}{\sqrt{2}\kappa} \left( k_{m+1} \otimes (1-m) P_1 + \bar k_{m+1} \otimes (1-m) P_2 \right) +\mathcal{O}\left(\frac{1}{\kappa^2}\right)\label{Deltak}
\end{align}
and
\begin{align}
\Delta (\bar k_m) = &   \mathbbm1 \otimes \bar k_m +\bar k_m \otimes \left( \mathbbm1 + m\frac{P_+}{\kappa} \right)  + \frac{m-1}{\sqrt{2}\kappa}M^{+-} \otimes  A_{1+m,1} \nonumber\\
&+ \frac{i}{\kappa} \left((M^{-1} \otimes \left(\frac{m+1}{2}A_{m1} + \frac{m-1}{2} A_{0,m+1} \right) - M^{-2} \otimes \left( \frac{m+1}{2} S_{m1} - \frac{m-1}{2}S_{0,m+1} \right)  \right) \nonumber \\
&+ \frac{i}{\sqrt{2}\kappa} \left( k_{m+1} \otimes (1-m) P_2 + \bar k_{m+1} \otimes (1-m) P_1 \right)+\mathcal{O}\left(\frac{1}{\kappa^2}\right).\label{Deltabk}
\end{align}

In the supertranslation sector we get
\begin{align}
\Delta ( A_{pq}) = &\mathbbm1 \otimes A_{pq} + A_{pq} \otimes  \left( \mathbbm1 +(p+q-1)\frac{P_+}{\kappa} \right)    \nonumber \\
 &+ \frac{i}{\sqrt{2}\kappa}(1-p) \left( A_{p+1, q} \otimes P_1 + S_{p+1, q} \otimes P_2 \right) \nonumber \\
 &+ \frac{i}{\sqrt{2}\kappa}(1-q) \left( A_{p, q+1} \otimes P_1 + S_{p, q+1} \otimes P_2 \right)+\mathcal{O}\left(\frac{1}{\kappa^2}\right)\label{DeltaS}
\end{align}
and
\begin{align}
\Delta ( S_{pq}) =& \mathbbm1 \otimes S_{pq} + S_{pq} \otimes  \left( \mathbbm1 + (p+q-1)\frac{P_+}{\kappa} \right)    \nonumber \\
& + \frac{i}{\sqrt{2}\kappa}(1-p) \left( A_{p+1, q} \otimes P_2 + S_{p+1, q} \otimes P_1 \right) \nonumber \\
& + \frac{i}{\sqrt{2}\kappa}(1-q) \left( A_{p, q+1} \otimes P_2 + S_{p, q+1} \otimes P_1 \right)+\mathcal{O}\left(\frac{1}{\kappa^2}\right).\label{DeltaA}
\end{align}

To obtain the antipodes of the deformed Hopf algebra it is often advantageous to make use of the defining property

\begin{align}
m \circ (S \otimes \mathbbm1) \circ \Delta (x) = \eta(x) \epsilon (x) = m \circ (  \mathbbm1 \otimes S) \circ \Delta (x).
\end{align}

Thus, one obtains
\begin{align}
S(k_m) = &  - k_m  \left( 1 + m\frac{P_+}{\kappa} \right)  - \frac{m-1}{\sqrt{2}\kappa}M^{+-} S_{1+m,1} \nonumber\\
&- \frac{i}{\kappa} \left((M^{-1}  \left(\frac{m+1}{2}S_{m1} + \frac{m-1}{2} S_{0,m+1} \right) - M^{-2}  \left( \frac{m+1}{2} A_{m1} - \frac{m-1}{2}A_{0,m+1} \right)  \right) \nonumber \\
&- \frac{i}{\sqrt{2}\kappa} \left( k_{m+1}  (1-m) P_1 + \bar k_{m+1}  (1-m) P_2 \right) +\mathcal{O}\left(\frac{1}{\kappa^2}\right),\label{Sk}\\
 S(\bar k_m) =&  - \bar k_m  \left( 1 + m\frac{P_+}{\kappa} \right)  - \frac{m-1}{\sqrt{2}\kappa}M^{+-} A_{1+m,1} \nonumber\\
&- \frac{i}{\kappa} \left((M^{-1}  \left(\frac{m+1}{2}A_{m1} + \frac{m-1}{2} A_{0,m+1} \right) - M^{-2}  \left( \frac{m+1}{2} S_{m1} - \frac{m-1}{2}S_{0,m+1} \right)  \right) \nonumber \\
&- \frac{i}{\sqrt{2}\kappa} \left( k_{m+1}  (1-m) P_2 + \bar k_{m+1}  (1-m) P_1 \right)+\mathcal{O}\left(\frac{1}{\kappa^2}\right)\label{Sbk}
\end{align}
and
\begin{align}
S ( A_{pq}) =&  -A_{pq} \left( 1 +(p+q-1)\frac{P_+}{\kappa} \right)    \nonumber \\
& - \frac{i}{\sqrt{2}\kappa}(1-p) \left( A_{p+1, q}  P_1 + S_{p+1, q} P_2 \right) \nonumber \\
 &- \frac{i}{\sqrt{2}\kappa}(1-q) \left( A_{p, q+1}  P_1 + S_{p, q+1}  P_2 \right)+\mathcal{O}\left(\frac{1}{\kappa^2}\right)\label{SA},\\
S ( S_{pq}) =& -S_{pq}  \left( 1 +(p+q-1)\frac{P_+}{\kappa} \right)    \nonumber \\
 &- \frac{i}{\sqrt{2}\kappa}(1-p) \left( A_{p+1, q} P_2 + S_{p+1, q} P_1 \right) \nonumber \\
 &- \frac{i}{\sqrt{2}\kappa}(1-q) \left( A_{p, q+1}  P_2 + S_{p, q+1}  P_1 \right)+\mathcal{O}\left(\frac{1}{\kappa^2}\right)\label{SS}.
\end{align}

The most important physical aspect of the co-product of an algebra is that it describes a rule of how to compose the associated charge for multiparticle systems. For example the co-product of $P_+$, $\Delta(P_+)$ tells us what is the total energy of a system of many particles, each having the energy $P_+^{(i)}$; in the case of just two particles, we have from \eqref{DeltaS}
\begin{equation}\label{DeltaP+}
  \Delta{P_+} = \mathbbm1\otimes P_+ + P_+\otimes\left( \mathbbm1+\frac{P_+}\kappa\right)+ O\left(\frac1{\kappa^2}\right)\,,
\end{equation}
so that for the two-particle system the total energy $P_+^{(1+2)}$ is
\begin{equation}\label{P+tot}
  P_+^{(1+2)} \equiv P_+^{(1)}\oplus P_+^{(2)} = P_+^{(1)}+ P_+^{(2)} + \frac1\kappa\, P_+^{(1)}\, P_+^{(2)} + O\left(\frac1{\kappa^2}\right)\,.
\end{equation}
Similarly the antipode tells you how to define the `inverse', denoted sometimes by $\ominus X \equiv S(X)$, so that $X\oplus (\ominus X)=0$. For example, the `negative momentum' in direction 1 is given by, \eqref{SS}
\begin{equation}\label{-P1}
  \ominus P_1= - P_1\left( 1+\frac{P_+}\kappa\right)\,.
\end{equation}

The presence of a  non-trivial Hopf structure, whose appearance is to be regarded as a quantum gravity effect, has important physical consequences. It may, for example, shed some new light on the ongoing discussion on black hole information paradox \cite{Hawking:1976ra}.  Recently, Hawking, Perry, and Strominger \cite{Hawking:2016msc}, \cite{Hawking:2016sgy} argued that the soft charges could dramatically change the standard analysis of black hole evaporation, because instead of being nearly bald (having - in the purely Einsteinian case - only mass and angular momentum as its charges) is in fact possessing a (potentially infinite) number of soft hair. Since the corresponding supertranslation charges are all conserved, their presence introduces previously overlooked correlations between early and late Hawking radiation, possibly rendering  the black hole evaporation process unitary. In fact, the hard and soft part are not conserved separately, but the total charge is:
\begin{align}
Q^-[f] &= Q^+[f],\quad Q^{\pm} = Q^{\pm}_s + Q^{\pm}_h,
\end{align}
where $\pm$ refers to past and future null infinity and $f$ is an arbitrary function on the sphere. Exchange of charge between soft and hard part during the time evolution would introduce the correlations between early and late quanta.

This argument was refuted in a series of papers \cite{Mirbabayi:2016axw}, \cite{Bousso:2017dny}, \cite{Bousso:2017rsx}, \cite{Javadinazhed:2018mle} where it is argued that soft hair could not resolve the black hole unitarity problem, because their time evolution is trivial and their conservation laws are automatically satisfied. The soft modes carrying them effectively decouple from the hard ones and moreover there is a canonical transformation which strips the hard modes from all the supertranslational charges. As a result the supertranslational charges evolve trivially and each charge stays the same all the way through from $\mathscr I^-$ to $\mathscr I^+$. One can say that although a black hole has infinitely many soft hair, they are perfectly combed, so that they do not tangle with the hard ones.

The presence of deformation changes this qualitative picture considerably, the reason being the presence of the co-product structure and the associate modification of the composition laws. Indeed consider eqs.\ \eqref{DeltaA} and \eqref{DeltaS}, describing the deformed coproduct. It follows that in the associate conservation laws the total supertranslation charge will depend on the momenta carried by hard particles. Indeed consider the total supertranslation charge of the incoming system of a soft particle, having zero Poincar\'e momentum and some  supertranslation charges $T_{pq}$ and a hard one carrying only Poincar\'e momenta $P_\mu$
\begin{align}
 T_{pq}^{Tot} =& T_{pq}   +\frac1\kappa\, (p+q-1)\, T_{pq}\, P_+   
 + \frac{i}{\sqrt{2}\kappa}(1-p) \left( T_{p+1, q}\,  P_1 + i T_{q, p+1}\,  P_2 \right) \nonumber \\
& + \frac{i}{\sqrt{2}\kappa}(1-q) \left( T_{p, q+1} \, P_1 + iT_{q+1, i} \, P_2 \right)+\mathcal{O}\left(\frac{1}{\kappa^2}\right).\label{TotalT}
\end{align}
If the coproduct was primitive the  $T_{pq}^{Tot}$ would be given purely by the eigenvalue $T_{pq}$ of the soft particle and likewise for the late quanta. This corresponds to the argument of \cite{Mirbabayi:2016axw, Bousso:2017dny, Bousso:2017rsx, Javadinazhed:2018mle}, because  the supertranslation charges conservation would constrain only soft particles. But in the case of $\kappa$-deformed BMS the total supertranslation charges of early and late quanta are also functions of the Poincar\'e momenta of the hard particles. This makes the clear separation of hard and soft part impossible. Notice that this effect is negligible at the LHC energy scale, since it is suppressed by $1/\kappa$, which is expected to correspond to the Planck scale.

Thus the fundamental property of Hopf algebras, the presence of a non-trivial coproduct structure, translates into an irreducible entanglement between hard momenta and supertranslational charges. This opens a possibility that early and late Hawking radiation quanta are correlated after all. One may say that as a result of deformation, possibly being associated with quantum gravity effects, the combed hair of black hole become tangled again. Of course much more detailed calculations are required to claim that the level of deformation induced entanglement is sufficient to solve the black hole information paradox, and we hope to address this in future publication.

Apart from these physical investigations our work suggests a number of possible more technical developments.  Firstly, it should be noticed  that the mathematical studies of
 coboundary Lie bialgebra structures on infinite-dimensional   Lie algebras and/or their supersymmetric extensions has   extensive literature, e.g. \cite{Taft,Michaelis,Ng}.
Particularly, the case of Jordanian deformation of the (super) Witt and Virasoro algebras have been considered in the literature before, see  \cite{Yang,Yuan} and reference therein. This suggests the possibility of studying  supersymmetric extensions of the BMS algebra and its deformation.

It was established recently \cite{Borowiec:2015wua} that the $\kappa$-Poincar\'e algebra with light-come deformation is associated with the integrable deformation of principal sigma model. This indicates that similar construction might be possible in the case of the $\kappa$-deformed BMS algebra. We will address this issue in a future publication.

More systematic study  of possible  real forms of obtained deformation  might be also of some interest.  The one, implicitly  assumed in the present paper, has been induced from the Poincar\'e reality condition: $l_m^\dagger=- \bar l_m, \ \bar l_m^\dagger=-l_m, \ T_{mn}^\dagger =-T_{nm}$.
It  manifests itself by the choice of Hermitian Poincar\'e generators making the twist (\ref{TwiEl}) $\dagger$- unitary.
More precisely, the BMS  algebra treated as a complex Lie algebra  can admit several real forms;  some of them are induced by the reality condition on the embedded  $\mathfrak{o}(4,\mathbb{C})$ Lie algebra
 \footnote{See  \cite{blt16,blt17b,blt17} for the discussion of  reality conditions on a complex Lie bialgebra as well as quantized enveloping algebra level in the case of  $\mathfrak{o}(4,\mathbb{C})$.}.

Another issue concerning  the choice of possible physical observables in a deformed algebra which might be better adjusted for  description of  relativistic kinematics in a quantum gravity regime. Twist deformation can offer a canonical solution to this problem (see e.g.  \cite{ABP,Toppan} and references therein).

\section*{Appendix}
In this section we will list the all-orders expressions for the coproducts and antipodes. The leading order of these expressions was discussed in the main text.

For the coproducts we have 
\begin{align}
\Delta (k_m) =&  \mathbbm1 \otimes k_m  + k_m \otimes \Pi_+^{m}
+ \frac{i(m-1)}{\sqrt{2}\kappa} k_0 \otimes  \Pi_+^{-1} S_{1+m,1} \nonumber\\
&- \frac{i}{\sqrt{2}\kappa} k_1 \otimes \left(\frac{m+1}{2}S_{m 1} + \frac{m-1}{2}S_{0,m+1}\right)
- \frac{i}{\sqrt{2}\kappa} \bar k_1 \otimes \left(\frac{m+1}{2}A_{m 1} - \frac{m-1}{2}A_{0, m+1}\right)  \nonumber\\
& - \frac{m-1}{(\sqrt{2}\kappa)^2}  \left(  k_1 \otimes \Pi_+^{-1} S_{01} S_{1,m+1} + \bar k_1 \otimes  \Pi_+^{-1}A_{01} S_{1+m,1}  \right) \nonumber \\
& + \sum_{n=1}^{\infty}  \left(\frac{-i}{\sqrt{2}\kappa}\right)^n{n+m-2\choose n} (k_{m+n} \otimes f_n^{}  \Pi_+^{m}+ \bar k_{m+n} \otimes \bar f_n^{}  \Pi_+^{m} )
\end{align}
and
\begin{align}
\Delta (\bar k_m) = &  \mathbbm1 \otimes \bar k_m+\bar k_m \otimes \Pi_+^{m}
+  \frac{i(m-1)}{\sqrt{2}\kappa}k_0 \otimes  \Pi_+^{-1} A_{1+m,1}\nonumber  \\
&- \frac{i}{\sqrt{2}\kappa} k_1 \otimes \left(\frac{m+1}{2}A_{m 1} + \frac{m-1}{2}A_{m+1,0}\right)
- \frac{i}{\sqrt{2}\kappa} \bar k_1 \otimes \left(\frac{m+1}{2}S_{m 1} - \frac{m-1}{2}S_{m+1, 0}\right) \nonumber \\
&  - \frac{m-1}{(\sqrt{2}\kappa)^2}  \left(  k_1 \otimes  \Pi_+^{-1} S_{01} A_{1,m+1} +\bar k_1 \otimes  \Pi_+^{-1} A_{01} A_{1+m,1}  \right) \nonumber \\
& + \sum_{n=1}^{\infty}  \left(\frac{-i}{\sqrt{2}\kappa}\right)^n{n+m-2\choose n} (k_{m+n} \otimes g_n^{} \Pi_+^{m}+ \bar k_{m+n} \otimes \bar g_n^{}  \Pi_+^{m} ),
\end{align}
where the elements $f_n, , g_n, \bar f_n, \bar g_n$ are defined by the following recurrent relations
\begin{align}
f_1^{} =\bar g_1&= S_{01}, \quad g_1=\bar f_1^{} =  A_{01}\,, \\
f_{n+1}^{} & = f_n^{} S_{01} - \bar f_n^{} A_{01}\,, \quad
\bar f_{n+1}^{}  = f_n^{} A_{01} +\bar f_n^{} S_{01}\,,\\
g_{n+1}^{} & = g_n^{} S_{01}  -\bar g_n^{} A_{01})\,, \quad
\bar g_{n+1}^{}  = g_n^{} A_{01} +\bar g_n^{} S_{01}\, .
\end{align}
 Here the usual  binomial notation ${n+m-2\choose n}=\frac{(n+m-2)(n+m-1)\ldots (m-1)}{n!}$ is in use and the infinite sums in the equations  (34), (35) become finite whenever $m\leq 1$.

In the supertranslation sector we get
\begin{align}
\Delta ( A_{pq}) =  \mathbbm1 \otimes A_{pq} + A_{pq} \otimes \Pi_+^{(p+q-1)}  \nonumber \\
 + \sum_{n = 1}^{\infty} \frac{1}{n!}\sum_{r,s}^{n \geq r+s >0}  \left(\frac{i}{\sqrt{2}\kappa}\right)^{r+s}\left(A_{p+r, q+s} \otimes g_{rs}^{(pq)} \Pi_+^{(p+q-1)}  + S_{p+r, q+s} \otimes f_{rs}^{(pq)}\Pi_+^{(p+q-1)}   \right)
\end{align}
and
\begin{align}
\Delta (S_{pq}) =  \mathbbm1 \otimes S_{pq} + S_{pq} \otimes \Pi_+^{(p+q-1)}   \nonumber \\
 + \sum_{n = 1}^{\infty} \frac{1}{n!} \sum_{r,s}^{n \geq r+s >0}  \left(\frac{i}{\sqrt{2}\kappa}\right)^{r+s}\left(A_{p+r, q+s} \otimes h_{rs}^{(pq)} \Pi_+^{(p+q-1)}  + S_{p+r, q+s} \otimes j_{rs}^{(pq)}\Pi_+^{(p+q-1)}   \right),
\end{align}
where we abbreviate
\begin{align}
g_{10}^{(pq)} =  (1-p)S_{01}, \quad g_{01}^{(pq)} =  (1-q) S_{01} ,\\
f_{10}^{(pq)} =  (1-p)A_{01}, \quad f_{01}^{(pq)} = (1-q)A_{01} ,\\
g_{r+1, s}^{(pq)} =  (1-(p+r))(g_{rs}^{(pq)} S_{01} + f_{rs}^{(pq)} A_{01}) ,\\
g_{r, s+1}^{(pq)} = (1-(q+s))(g_{rs}^{(pq)} S_{01} - f_{rs}^{(pq)} A_{01}) ,\\
f_{r+1, s}^{(pq)} =  (1-(p+r))(g_{rs}^{(pq)} A_{01} + f_{rs}^{(pq)} S_{01}) ,\\
f_{r, s+1}^{(pq)} =  (1-(q+s))(g_{rs}^{(pq)} A_{01}+ f_{rs}^{(pq)} S_{01}) ,
\end{align}
and
\begin{align}
j_{10}^{(pq)} = S_{01}, \quad j_{01}^{(pq)} = (1-q) S_{01} ,\\
h_{10}^{(pq)} =  (1-p)A_{01}, \quad h_{01}^{(pq)} = (1-q)A_{01} ,\\
j_{r+1, s}^{(pq)} = (1-(p+r))(j_{rs}^{(pq)} S_{01} + h_{rs}^{(pq)} A_{01}) ,\\
j_{r, s+1}^{(pq)} = (1-(q+s))(j_{rs}^{(pq)} S_{01} + h_{rs}^{(pq)} A_{01}) ,\\
h_{r+1, s}^{(pq)} = (1-(p+r))(j_{rs}^{(pq)} A_{01} + h_{rs}^{(pq)} S_{01}) ,\\
h_{r, s+1}^{(pq)} = (1-(q+s))(-j_{rs}^{(pq)} A_{01}+ h_{rs}^{(pq)} S_{01}).
\end{align}
Let us note that for the important Poincar\'{e} subalgebra these results coincide with those found in \cite{Borowiec:2014aqa}.
One can also notice that the first lines in the formulas (34), (35), (42), (43) represent the  coproduct  deformed by the purely Jordanian twist $\mathcal{J}=\exp\left( - k_0 \otimes \sum_{j=0}^{\infty} \frac{1}{j+1} \left(\frac{-i S_{11}}{\sqrt{2}\kappa}\right)^{j+1}\right)$ while the remaining lines come from the twist extension (c.f. (\ref{TwiEl})).
 \newline
For the antipodes it is in general only possible to write down implicit formulas like
\begin{align}
S ( A_{pq}) =&- A_{pq}  S(\Pi_+^{ -(1-(p+q))})  \nonumber \\
& - \sum_{n = 1}^{\infty} \sum_{r,s}^{n \geq r+s >0} \frac{1}{n!} \left(A_{p+r, q+s}  S(g_{rs}^{(pq)}) S(\Pi_+^{ -(1-(p+q))}) + S_{p+r, q+s} S(f_{rs}^{(pq)}) S(\Pi_+^{ -(1-(p+q))})  \right), \\
S ( S_{pq}) =&- S_{pq}  S(\Pi_+^{- (1-(p+q))})  \nonumber \\
& - \sum_{n = 1}^{\infty} \sum_{r,s}^{n \geq r+s >0} \frac{1}{n!} \left(A_{p+r, q+s}  S(h_{rs}^{(pq)}) S(\Pi_+^{ -(1-(p+q))}) + S_{p+r, q+s} S(j_{rs}^{(pq)}) S(\Pi_+^{- (1-(p+q))})  \right),
\end{align}
and similar for the superrotations.
\newline

To define the $q$-analog of the Hopf algebra at hand (see \cite{Borowiec:2014aqa}) one also needs to make sure that all structures including the coproduct are at most polynomial in the generators. In the Poincare sub-Hopf algebra this is achieved with the help of $\Pi_+$ and $\Pi_+^{-1}$. One has to define all the Hopf algebra structures including these auxiliary objects which replace $S_{1\,1}=-i\sqrt{2}P_{+}$. The only nontrivial commutators are
\begin{align}
[ k_m, \Pi_+] = & \frac{ (m-1)}{\sqrt{2}\kappa} S_{1+m, 1}, \\
 [ \bar k_m, \Pi_+] = & \frac{ (m-1)}{\sqrt{2}\kappa} A_{1+m, 1}
\end{align}
and
\begin{align}
0 =& [ k_m, \Pi_+ \Pi_+^{-1}] = \Pi_+ [k_m, \Pi_+^{-1}] + [k_m, \Pi_+] \Pi_+^{-1} \\
\Rightarrow [k_m, \Pi_+^{-1}] =& \frac{-(m-1)}{\sqrt{2}\kappa}S_{1+m, 1} (\Pi_+^{-1})^2 ,\\
 [\bar k_m, \Pi_+^{-1}] =& \frac{-(m-1)}{\sqrt{2}\kappa}A_{1+m, 1} (\Pi_+^{-1})^2.
\end{align}
The coproducts then take the form
\begin{align}
\Delta (\Pi_+) &= \mathbbm1 \otimes \mathbbm1 + \frac{i}{\sqrt{2}\kappa} \Delta (S_{11}) \\
&= \Pi_+ \otimes \Pi_+, \\
\Delta (\Pi_+^{-1}) &= \Pi_+^{-1} \otimes \Pi_+^{-1}
\end{align}
and for the counits and antipodes one has
\begin{align}
\epsilon(\Pi_+) &= \epsilon(\Pi_+^{-1}) = 1, \\
S(\Pi_+) &= \Pi_+^{-1}, \quad S(\Pi_+^{-1}) = \Pi_+.
\end{align}
Note that for general generators of the BMS algebra (apart from the purely Jordanian case) this is not fulfilled in the current notation, e.g. the coproducts for $k_m$, $\bar k_m$ contain infinite power series (in $1/\kappa$) for $m>1$. Finding a reformulation to remedy the situation, if possible, will be the goal of future research.

\subsection*{Acknowledgments}
This research was supported by Polish National Science Center (NCN), project UMO-
2017/27/B/ST2/01902.  We thank Michele Arzano for discussion on the physical role of Hopf structures in the context of BMS symmetry.

\end{document}